\newcommand {\ed}[1]{{\color{black}{#1}}}
\documentclass[sigconf]{acmart}

\copyrightyear{2024}
\acmYear{2024}
\setcopyright{rightsretained}
\acmConference[ASSETS '24]{The 26th International ACM SIGACCESS Conference
on Computers and Accessibility}{October 27--30, 2024}{St. John's, NL, Canada}
\acmBooktitle{The 26th International ACM SIGACCESS Conference on Computers
and Accessibility (ASSETS '24), October 27--30, 2024, St. John's, NL, Canada}
\acmDOI{10.1145/3663548.3675651}
\acmISBN{979-8-4007-0677-6/24/10}

\usepackage{hyperref}
\usepackage{balance}

\begin{document}

\title[What Intersectional, Neurodivergent Lived Experiences Bring to Accessibility Research]{``I Am Human, Just Like You'': What Intersectional, Neurodivergent Lived Experiences Bring to Accessibility Research}

\author{Lindy Le}
\email{lindle@microsoft.com}
\affiliation{%
  \institution{Microsoft}
  \city{Redmond}
  \state{Washington}
  \country{USA}
}

\renewcommand{\shortauthors}{Lindy Le}

\begin{abstract}
The increasing prevalence of neurodivergence has led society to give greater recognition to the importance of neurodiversity. Yet societal perceptions of neurodivergence continue to be predominantly negative. Drawing on Critical Disability Studies, accessibility researchers have demonstrated how neuronormative assumptions dominate HCI. Despite their guidance, neurodivergent and disabled individuals are still marginalized in technology research. In particular, intersectional identities remain largely absent from HCI neurodivergence research. In this paper, I share my perspective as an outsider of the academic research community: I use critical autoethnography to analyze my experiences of coming to understand, accept, and value my neurodivergence within systems of power, privilege, and oppression. Using Data Feminism as an accessible and practical guide to intersectionality, I derive three tenets for reconceptualizing neurodivergence to be more inclusive of intersectional experiences: (1) neurodivergence is a functional difference, not a deficit; (2) neurodivergent disability is a moment of friction, not a static label; and (3) neurodivergence accessibility is a collaborative practice, not a one-sided solution. Then, I discuss the tenets in the context of existing HCI research, applying the same intersectional lens. Finally, I offer three suggestions for how accessibility research can apply these tenets in future work, to bridge the gap between accessibility theory and practice in HCI neurodivergence research.
\end{abstract}

\ccsdesc[500]{Social and professional topics~People with disabilities}
\ccsdesc[500]{Human-centered computing~Accessibility theory, concepts and paradigms}

\keywords{neurodivergence, intersectionality, Autism, ADHD, neurodiversity, critical theory, disabilities studies, intersectional HCI}

\maketitle

\section{Introduction}
Within the past few years, the subject of neurodiversity has gained increased attention in many public spheres, permeating domains such as employment~\cite{lefevre2023neurodiversity, krzeminska2019advantages}, healthcare~\cite{wright2022understanding, izuno2023neurodiversity}, education~\cite{mirfin2020responding, baker2023pathology}, and social media~\cite{jawed2023digital, johnson2021normalizing}. Recent research explains why society's active focus on neurodiversity is important---approximately 15-20\% of the world's population is neurodivergent~\cite{doyle2020neurodiversity}. Despite the prevalence and raised awareness of neurodiversity, global perceptions of neurodivergence continue to be predominantly negative~\cite{awosanya2021public, ha2014living, araujo2023autism, mueller2012stigma}, contributing to systemic harms that affect individuals' physical and mental health~\cite{turnock2022understanding, fuermaier2012measurement, lebowitz2016stigmatization}. These stigmatizing attitudes have also been internalized by neurodivergent individuals themselves, exacerbating the discrimination, isolation, and rejection caused by society~\cite{botha2022autism, han2022systematic, masuch2019internalized}. It was in response to these experiences of marginalization~\cite{kapp2020autistic, walker2021toward} that the neurodiversity movement was formed, by neurodivergent individuals\footnote{Throughout this paper, I use identity-first versus person-first language, because that is my personal preference. However, it is important to note that disability-related language preferences vary between individuals~\cite{kenny2016terms, dwyer2022stigma, taboas2023preferences, buijsman2023autistic}, and conclusions should not be drawn about the disability-related language preferences of other neurodivergent or disabled individuals.} advocating for their rights.

Similarly, Human-Computer Interaction (HCI) has witnessed mounting interest in neurodivergence research~\cite{williams2020perseverations}, particularly, social skills interventions for autistic children~\cite{spiel2019agency}. Yet many of these technologies are predicated on normative assumptions that problematize neurodivergent bodies and minds~\cite{spiel2022adhd, williams2023counterventions}. Influenced by Critical Disability Studies, neurodivergent and disabled scholars have critiqued the involvement, or lack thereof, of people with disabilities in existing HCI accessibility practice~\cite{mack2021we, spiel2021purpose, mankoff2010disability, bennett2019promise, williams2023counterventions}. In spite of their contributions, many of these authors note, as recently as last year's ASSETS conference~\cite{mcdonnell2023tackling}, that the wider research community continues to produce technology that harms neurodivergent and disabled users~\cite{gadiraju2023wouldn, hundt2024love}. 

This paper builds on prior work to bridge the gap between accessibility theory and practice in neurodivergence research, using critical autoethnography to surface everyday experiences of intersectional neurodivergence, which are marginalized by the academic research community. Critical autoethnography follows the approach of traditional autoethnography while ``bring[ing] attention to the ways cultures are created and compromised through institutional, political, social, and interpersonal relations of power''~\cite{holman2018creative}. To stay true to the nature of critical autoethnography, the majority of this paper is written in first-person~\cite{poulos2021essentials}, with self-reflexivity woven throughout the text~\cite{boylorn2020critical}. 

I constructed the autoethnographic narratives using data from my journey of discovering, processing, and coming to terms with my neurodivergent diagnoses\footnote{The discussion of neurodivergence in this paper is largely framed within my experiences of autism and ADHD (although I do make references to C-PTSD and mental health conditions), as the history of neurodiversity has meant that research and conversations about neurodivergence are dominated by autistic and ADHD experiences. I encourage readers to consider other experiences of neurodivergence that are missing from the conversation.} (Section \ref{sec:reflection-questions}). In the same way that accessibility research prioritizes lived experiences (e.g., ~\cite{mankoff2010disability, spiel2020nothing, mack2022anticipate, williams2023counterventions, mack2021we}), critical autoethnography ``foregrounds embodiment as an important and valuable source of sense making''~\cite{holman2018creative}, promoting relational connection between the author and reader via stories that are emotionally vulnerable and rich in complexity~\cite{holman2018creative}. Thus, I present the narratives raw in their entirety (Section \ref{sec:narratives}), before analyzing them \ed{through an intersectional lens (Section \ref{sec:findings}), as part of my critical autoethnographic approach (Section \ref{sec:critical-autoethnography}). To perform this analysis, I used the principles outlined in Data Feminism---a community-reviewed, broadly accessible, and intersectional framework for working with data---developed by D'Ignazio and Klein~\cite{d2023data}.} 

From this analysis, I derive three tenets for a more holistic and compassionate conceptualization of neurodivergence: 1) neurodivergence is a functional difference that can be adaptive, depending on context; 2) neurodivergent disability is a moment of friction enacted by an individual's abilities and their environment; and 3) neurodivergence accessibility learns from neurodivergent traits to mutually benefit both neurotypical and neurodivergent people. By framing neurodivergence as a functional difference, I am shifting the conversation away from debates between the medical and social models of disability, which have divided both the accessibility research~\cite{williams2019prefigurative, kirkham2021disability, spiel2020nothing} and broader neurodivergent~\cite{Ashburn_Edwards_Onaiwu_McConnell_2023, bagatell2010cure, kapp2020autistic} communities. Moreover, by centering the tenets around everyday lived experiences and Data Feminism's practical approach to intersectionality, I am moving the discussion beyond Critical Theory---which has been critiqued for its inaccessibility, impracticality, and ``orient[ation] more towards the academy than towards activism''~\cite{shakespeare2013disability}---to pragmatic applications.
 
As an outsider of the academic research community, I provide a perspective that has often been acknowledged as essential to accessibility research~\cite{hamraie2019crip, hsueh2023cripping, williams2023cyborg}, but largely absent from HCI~\cite{harrington2023working, sum2022dreaming, chordia2024social}: everyday lived experiences of disability as it intersects with identities such as race, class, gender, and sexuality. At the same time, my experiences are my own, and I do not claim to speak for anyone else in the disability or neurodivergent communities. Although I share my experiences of marginalization, I also acknowledge the privilege I have to be able to voice these experiences to begin with (Section \ref{sec:limitations}). Thus, while researchers are the primary target audience of this paper, my hope is that this paper empowers other underrepresented voices to share their experiences as well. My aim is to foster an understanding of neurodivergence that recognizes the complexity and humanity of neurodivergent individuals. For all the ways that we may be different, we are the same in that we are all human---and history has shown us that adaptive societies ``thrive on inclusion and diversity,'' where members ``collaborate with the most diverse set of other humans, actively learning from everyone''~\cite{gopin2022compassionate}.

To start, I present a brief overview of neurodiversity (Section \ref{sec:neurodiversity}) and explore how societal perceptions of neurodivergence have shaped accessibility research and HCI (Section \ref{sec:accessibility-HCI}). After introducing core concepts in intersectionality and Data Feminism (Section \ref{sec:intersectionality}), I explain critical autoethnography as my method (Section \ref{sec:critical-autoethnography}). I then offer personal narratives that contextualize the problems with prevailing attitudes toward neurodivergence in everyday lived experiences of intersectional disability (Section \ref{sec:narratives}). From an analysis of these narratives, I derive three tenets for reconceptualizing neurodivergence (Section \ref{sec:findings}). I then analyze these tenets in the context of HCI, continuing to use Data Feminism as an intersectional lens (Section \ref{sec:discussion}). Finally, I provide recommendations for practicing these neurodivergence tenets in future technology research (Section \ref{sec:implications}). The main contributions of this paper are 1) three tenets for a more inclusive and intersectional understanding of neurodivergence, 2) an intersectional analysis of these tenets in existing HCI research, and 3) suggestions for engaging with these tenets in future accessibility research.

\section{Background and Related Work}
In this section, I define neurodiversity and related terms, situating them within societal attitudes toward disability (Section \ref{sec:neurodiversity}); review how accessibility research and HCI has engaged with these understandings of neurodivergence (Section \ref{sec:accessibility-HCI}); and introduce the intersectional principles from Data Feminism (Section \ref{sec:intersectionality}).

\subsection{Overview of the Neurodiversity Movement} \label{sec:neurodiversity}
The concept of neurodiversity emerged during the autism rights movement\footnote{Though commonly attributed to Australian sociologist Judy Singer, neurodiversity was recently brought forward to have  originated collectively from online activist communities~\cite{botha2024neurodiversity}.} in the late 1990s~\cite{walker2021toward}. \textit{Neurodiversity} refers to ``neurological diversity''~\cite{singer1999can}, or ``the diversity of human brains and minds – the infinite variations in neurocognitive functioning within our species''~\cite{walker2021neurodiversity}. Within neurodiversity, individuals whose brains ``diverge from dominant cultural standards of neurocognitive functioning'' (e.g., autistic, dyslexic, epileptic)\footnote{``The label neurodivergent includes psychiatric diagnoses, learning disabilities, brain injuries, ADHD, and cognitive disabilities of all sorts---any brain that doesn't think in `conventional' or expected or `neurotypical' ways''~\cite{shew2023against}.} are described as \textit{neurodivergent}, a term coined by autistic activist Kassiane Asasumasu~\cite{walker2021toward}. Conversely, individuals whose minds conform to societal standards are referred to as \textit{neurotypical}\footnote{Neurotypical is not a synonym for non-autistic, in the same way that neurodivergent is not a synonym for autistic~\cite{walker2021neurodiversity}. An equivalent term for non-autistic is \textit{allistic}~\cite{monk2022use}.}''~\cite{walker2021neurodiversity}.

The neurodiversity movement arose in objection to the medical model of disability, which sees neurodivergence as an impairment in the individual, a disease to be ``cured''~\cite{walker2021toward}. In response to dehumanizing treatment and prejudice, neurodiversity advocates adopted the social model of disability, which attributes the primary cause of disability to ``how society fails to accommodate and accept impaired individuals''~\cite{chapman2020neurodiversity}. Instead of eliminating neurodivergence, these activists focus on removing societal barriers and celebrate their neurodivergence as unique traits~\cite{kapp2020autistic}.

The conceptual divide between the medical and social models of disability has led to disagreements between academics. For while the medical model of disability has been criticized for its overly negative perception of neurodivergence, the social model of disability has also been criticized for an overly positive outlook that ignores how bodily manifestations of neurodivergence can be painful and inherently disabling, regardless of societal accommodations~\cite{chapman2020neurodiversity, anastasiou2013social}. As a result of these debates, researchers have explored other theoretical approaches to neurodivergence (e.g.,~\cite{botha2021critical, davidson2010cuts, sarrett2012autistic, kapp2013interactions, anderson2022autism, kintzinger2021equity, chapman2021neurodiversity, maiese2021enactivist, nevsic2023ecological}).

However, the social model/medical model binary is a false dichotomy:  In fact, neurodiversity advocates have argued that the neurodiversity paradigm allows for an understanding of disability that ``acknowledge[s] the transaction between inherent
weaknesses and the social environment''~\cite{kapp2020autistic}, one that ``transcend[s] the construction of binary oppositions such as `Medical Model vs. Social Model'\,''~\cite{Singer_2017}. Despite arguments that the social model is a complement, not replacement for the medical model~\cite{kapp2020autistic, oliver2013social}, society continues to conflate the neurodiversity movement's protest against ``curing'' autism with the idea that a social model understanding of neurodivergence implies no place for medicine or biology~\cite{ortega2009cerebral, kapp2020autistic}.  

While this paper acknowledges that Critical Theory contributions have been immensely valuable in ``opening up cultural representations and discourses to critical examination''~\cite{shakespeare2013disability}, it also aligns with work from neurodivergent and disabled scholars that has described these models of disability as ``academic-political tool[s]''~\cite{kapp2020autistic} that are difficult to operationalize in practice by non-academics~\cite{shakespeare2013disability}. Therefore, rather than further these theoretical disputes, this research focuses on bridging perspectives on neurodivergence, as prior work has suggested that there is substantial overlap between different conceptualizations of neurodivergence outside of academia~\cite{kapp2013deficit}.

\subsection{Neurodivergence, Accessibility \&  HCI} \label{sec:accessibility-HCI}
The social model/medical model binary's influence also extends to accessibility research~\cite{mankoff2010disability}, with clinical perspectives continuing to dominate HCI research on neurodivergence, from wearables~\cite{williams2020perseverations} to games~\cite{spiel2021purpose}. More recently, neurodivergent HCI researchers have highlighted how technology for ADHD users has largely framed them as ``\,`problems to solve'\,'' and ignored the needs, wants, and interests of the target population, even in the face of participant resistance~\cite{spiel2022adhd}. Although neurodiversity has been introduced in HCI~\cite{dalton2013neurodiversity}, neurodivergence is still regarded primarily as a negative experience that can be made better with technology. For example, Competency-Based Design describes competencies as participants' ability to interact with apps (e.g., browse YouTube) \textit{in spite of their intellectual disability} ~\cite{bayor2021toward}, as opposed to discussing the strengths inherent in participants' natural behaviors. 

Recent work has engaged in co-designing with neurodivergent individuals throughout the entire research process, empowering participants to express their needs and desires~\cite{stefanidi2023children, santos2022co, ravn2022co}. Yet empowerment is often defined based on whether neurodivergent users can behave in a way that is consistent with societal expectations. In Stefanidi et al.'s study, empowerment is defined by therapists and educators, who ``link the act of reflecting with empowering children [with ADHD] to express themselves in this way via technology''~\cite{stefanidi2023children}. But what if these children are already reflecting via other behaviors that we don't recognize, because we expect reflection to look a certain way? And even if communication is necessary between parents, children, and therapists, why is the onus on the children to adapt their behavior versus others in their care system? Nonverbal autistic writer Mel Baggs makes a powerful statement in their video ``In My Language'' about how intelligence and personhood are judged by whether one can express themselves in a way that is recognized as ``thought'' or ``language'' by neurotypical society~\cite{Baggs_2007}.

Researchers are beginning to push back on the notion that neurodivergent and disabled people can only be beneficiaries of technology. Williams et al. define ``counterventions'' as ``intervention[s] for interventionists,'' which locate problems of disability in social norms, as opposed to the disabled person~\cite{williams2023counterventions}. Likewise, Hundt et al. planned to design a ``social robot for NSD [Neurotypical Spectrum Disorder]'' but terminated the project based on the risk of harm to neurotypical people~\cite{hundt2024love}. Researchers are also starting to depict more holistic and complex portrayals of neurodivergent individuals by exploring their experiences of disability in the everyday~\cite{williams2023cyborg, kender2023banal, mok2023experiences}. Last year, Boyd introduced ``Celebratory Technologies for Neurodiversity'' to promote positive stories of neurodivergence and reduce stigma~\cite{boyd2023conceptualizing}. This paper contributes to this growing body of work (including scholarship in fields adjacent to HCI, e.g., neuro-positionality~\cite{stark2023disabling} and neuroqueer technoscience~\cite{rauchberg2022imagining}) by examining lived experiences of neurodivergence through an intersectional lens (Section \ref{sec:findings})---a topic that has gained recent attention in HCI accessibility practice~\cite{harrington2023working, sum2022dreaming} but is largely missing from neurodivergence research~\cite{Srinivasan_2023, ramirez2024adult, eagle2023you}---and providing recommendations on how researchers can be more inclusive of intersectional, neurodivergent identities (Sections \ref{sec:discussion} \& \ref{sec:implications}).

To clarify, this paper does not condemn medicalized approaches to neurodivergence research, as the ability to theorize about neurodiversity without acknowledging the sometimes harsh realities of neurodivergent disability is a privilege. Nor does it promote a purely positive perspective on neurodivergence, as this reductionist view risks becoming ``inspiration porn''~\cite{grue2016problem} or disability denial. Instead, this research seeks to bridge understandings of neurodivergence and disability in accessibility theory and practice, in the hopes of encouraging researchers to design technology that recognizes the complexity of neurodivergent individuals as humans, not all that different from themselves.

\subsection{Core Concepts in Intersectionality and Data Feminism} \label{sec:intersectionality}
In recognition of the ways HCI has marginalized the history of Black women in practicing intersectionality~\cite{rankin2019straighten}, this section begins with Collins and Bilge's definition of \textit{intersectionality}:

\begin{quote}
``Intersectionality is a way of understanding and analyzing complexity in
the world, in people, and in human experiences. The events and conditions
of social and political life and the self can seldom be understood as shaped
by one factor. They are shaped by many factors in diverse and mutually
influencing ways. When it comes to social inequality, people’s lives and the
organization of power in a given society are better understood as being
shaped not by a single axis of social division, be it race or gender or class,
but by many axes that work together and influence each other.''~\cite{collins2020intersectionality}
\end{quote}

Similar to the gap between accessibility theory and practice~\cite{mcdonnell2023tackling}, researchers outside of HCI have acknowledged ``the growing divide between cultural studies scholarship which theorizes, and methodologically-driven research which deploys the construct in practice''~\cite{rice2019doing}. Thus, this paper contextualizes intersectionality within the Data Feminism framework articulated by D'Ignazio and Klein~\cite{d2023data} (Appendix \ref{data-feminism-principles}), because it has been openly reviewed by the community and is targeted toward multiple audiences, including scholars across the sciences and humanities, industry professionals, and activists~\cite{D’Ignazio_Klein}. The aim is to present an accessible and practical application of intersectionality within technology research, advancing prior work on integrating intersectionality into HCI~\cite{rankin2019straighten, harrington2023working, erete2023method, schlesinger2017intersectional, chordia2024social}. 

D'Ignazio and Klein describe Data Feminism as ``a way of thinking about data, both their uses and their limits, that is informed by direct experience, by commitment to action, and by intersectional feminist thought''~\cite{d2023data}. Integral to understanding data feminism are the concepts of power, privilege, oppression, and co-liberation. \textit{Power} refers to ``the current configuration of structural privilege and structural oppression, in which some groups experienced unearned advantages [\textit{privilege}]---because various systems have been designed by people like them and work for people like them---and other groups experience systematic disadvantages [\textit{oppression}]---because those same systems were not designed by them or with people like them in mind''~\cite{d2023data}. They describe the goal of data feminism not as ``fairness'' but \textit{co-liberation}: ``mutual benefit, from members of both dominant groups and minoritized groups''~\cite{d2023data}. \label{co-liberation}

\section{Methodology}

\ed{The aim of this paper is to reconceptualize neurodivergence to be more inclusive of intersectional experiences. To this end,} I first introduce, explain, and justify critical autoethnography as a method in HCI (Section \ref{sec:critical-autoethnography}). Then, I detail my process for conducting critical autoethnography, describing how I constructed my personal narratives (Section \ref{sec:reflection-questions}).

\subsection{Critical Autoethnography} \label{sec:critical-autoethnography}
Autoethnography has been rising in popularity within HCI~\cite{rapp2018autoethnography, desjardins2021introduction}, particularly in the study of how race~\cite{erete2023method, erete2021can}, gender~\cite{spiel2021they, kaltenhauser2024playing}, disability~\cite{mack2022chronically, fussenegger2022depending, jain2020navigating}, and other marginalized identities affect technology use or participation in academia. Poulos describes \textit{autoethnography} as ``an observational, participatory, and reflexive research method that uses writing about the self in contact with others to illuminate the many layers of human social, emotional, theoretical, political, and cultural praxis (i.e., action, performance, accomplishment)''~\cite{poulos2021essentials}. Autoethnography has been recognized as an accessible approach for conducting research, especially for those who face ``physical and/or cultural obstacles associated with a disability''~\cite{polczyk2012autoethnography}. While researchers have responded to concerns about autoethnography's potential self-indulgence and lack of rigor by engaging in collaborative autoethnography~\cite{lapadat2017ethics, mack2022chronically, jain2020navigating}, such methods can be inaccessible for those writing from outside the academic research community. 

Instead, I turn to critical autoethnography as a way of attending to considerations of rigor and a multiplicity of perspectives, by ``highlight[ing] intersectional experiences of marginalization and interrogat[ing] social injustices''~\cite{boylorn2020critical}. Key to this process is \textit{reflexivity}, which ``require[s] researchers to acknowledge the inevitable privileges we experience alongside marginalization and to take responsibility for our subjective lenses''~\cite{boylorn2020critical}. Holman defines three goals of \textit{critical autoethnography}: (1) ``to examine systems, institutions, and discourses that privilege some people and marginalize others;'' (2) ``to mobilize and develop the explanatory frameworks that critical theory provides us---by putting theory into action through storytelling;'' and (3) ``to build new knowledge about the social world in order to stimulate new practices''~\cite{holman2018creative}. In addition to being an accessible research method, critical autoethnography also has the benefit of making theoretical research more accessible to applied practitioners, as it ``brings critical theory---a philosophical approach to culture that interrogates ideology, power, and the structural constraints of actors in a cultural system (e.g., oppression)---into direct conversation with audiences through autoethnographic texts''~\cite{poulos2021essentials}. \ed{I therefore analyzed my autoethnographic narratives using Data Feminism, because it is an intersectionality framework that operationalizes Critical Theory in a way that is practical and accessible to a broad audience~\cite{d2023data}.}

\subsection{Reflection Questions} \label{sec:reflection-questions}
I constructed my personal narratives around my experiences of getting late-diagnosed with autism and ADHD, including the process of coming to understand, accept, and value my neurodivergence. From spring of 2023 to spring of 2024, I had captured my thoughts in various notebooks, both paper and digital; on copies of books and research papers that I read; and over email or text message exchanges.  As part of my data gathering efforts, I also reviewed documentation from my therapist. Afterwards, I identified the following reflection questions\footnote{\ed{I use the label ``reflection questions'' in lieu of the more traditional ``research questions'' to distinguish this process from my later application of the tenets to accessibility research (Sections \ref{sec:discussion} \& \ref{sec:implications}), though the idea is similar---as seen by the parenthetical remarks in RQ1, RQ2, and RQ3.}} as common themes throughout the data:  

\begin{description}
    \item[\textbf{RQ1}] {How can I (/accessibility research) reconcile my experience of neurodivergence with the diverse experiences of neurodivergence across different individuals?}
    \item[\textbf{RQ2}] {How can I (/accessibility research) affirm my neurodivergent strengths while supporting my limitations?}  
    \item[\textbf{RQ3}] {How can I (/accessibility research) contribute to a society where all neurodivergent individuals are not only accepted but valued?} 
\end{description}

Then, I enumerated a list of all stories contained within the data, drawing connections between the stories and the reflection questions. I reflected on my experiences of marginalization and privilege across each story, using the principles defined in Data Feminism as a guide for operationalizing intersectional theory~\cite{d2023data} (Appendix \ref{data-feminism-principles}). For example, I followed ``Principle 2: Challenge power'' and ``Principle 4: Rethink binaries and hierarchies'' by describing the ways I demonstrate strength, creativity, and agency in navigating my experiences of marginalization, resisting deficit narratives about neurodivergent and disabled people that imply they need ``saving''~\cite{d2023data}. Over many iterations, I removed and consolidated stories from the list to obtain a minimum yet representative set, ending up with the five narratives presented in the next section.  

\section{Personal Narratives} \label{sec:narratives}

Though the stories contained within these narratives are all true, the narratives have been constructed with examinations of power, privilege, and oppression as the primary goal, not historical accuracy. In contrast to other first person perspectives on neurodivergence in HCI~\cite{kender2023banal, williams2023cyborg, guberman2023actuallyautistic, spiel2022adhd}, this paper foregrounds the humanity of neurodivergent individuals, independent of technological context. Thus, I present the narratives raw in their entirety first, before using Data Feminism to analyze them in relation to the reflection questions (Section \ref{sec:findings}) and discussing their implications for accessibility research (Sections \ref{sec:discussion} \& \ref{sec:implications}). While this decision adds to the length of the paper, placing narratives in the Appendix would go against the nature of critical autoethnography as a relational practice~\cite{holman2018creative}, negating its power to make critical theory more accessible to readers (Section \ref{sec:critical-autoethnography}).

Despite my efforts to portray a more holistic picture of neurodivergent lived experiences, stories can only provide a glimpse into the complexity of the people described within. I have tried to stay true to the nature of autoethnography and write narratives that are ``compelling, striking, and evocative (showing or bringing forth strong images, memories, or feelings)''~\cite{poulos2021essentials}. This meant that sometimes, the stories are particularly raw and vulnerable, with the hope that while readers may not be able to picture themselves in the exact circumstances, they can connect with the very human feelings the narratives depict---stories of uncertainty, struggle, hope, joy, and pain~\cite{brown2022gifts}. Although the stories follow a narrative arc, I want to emphasize that reality is not so linear, and there are days where I still struggle to accept and value my neurodivergence.

\ed{Lastly, as a content note, the narratives include descriptions of mental health conditions (Narratives \ref{narrative-1} \& \ref{narrative-2}), an autistic meltdown (Narrative \ref{narrative-4}), and police violence against Black and Brown bodies (Narrative \ref{narrative-4}).}

\subsection{Narrative 1: Recognizing Intersectional Oppression in My History of Mis-/Missed Diagnoses} \label{narrative-1}
I am sitting in my therapist's virtual office, and I am having a bad day. In fact, I have been having so many bad days that my therapist has written ``Major Depressive Disorder'' at the top of their notes. 

After an awkward silence, they gently ask me how I've been feeling. Sighing, I tell them that nothing has changed, that every day is a losing battle against exhaustion. Tiredness has engraved itself so deeply into my bones that I cannot remember what pleasure and hope feel like anymore. I have no energy left for work, friends, or hobbies, because even getting out of bed has become a chore.

Rationally, I shouldn't be feeling this way. After all, I have no work obligations while on disability leave, I have a supportive group of people nearby, and I have multiple outlets for my personal interests. I even have perfectly sunny weather outside, for goodness' sake!

``The worst part is that I \textit{want} to be doing things with my life, and I have spent hours working on myself, improving my sleep, eating, and exercise habits.'' My face crumples and tears escape against my will. ``But none of that matters when I am constantly overwhelmed by my senses and emotions. I can't bear to leave my house because the sound of birds grates on my ears, the smell of savory food churns my stomach, and the sight of pedestrian crossings hurts my head. And that's not to mention how anxiety-inducing it is for me to talk to people. In conversations, half the time I spend deciphering the meaning behind certain phrases and mannerisms, and the other half I spend planning my words and their delivery to avoid being misconstrued.''

My therapist watches as I cover my eyes with my fists and curl into myself. I think about how disappointed my family becomes when I decline yet another social gathering, and shame tightens its grip on my chest, piercing my heart with an all-consuming belief: \textit{I am irreparably broken and therefore, unlovable}. 

``Did you know that autism is often misdiagnosed as anxiety and depression in women?~\cite{kirkovski2013review, gesi2021gender, dell2023misdiagnoses}''

I look up in surprise and confusion. They continue, ``Your fatigue, sensory processing issues, and social-emotional struggles could be indicative of autistic burnout. Autistic burnout can result from the chronic stress of conforming to neurotypical expectations.~\cite{raymaker2020having, mantzalas2022autistic, higgins2021defining}''

``Does this mean that I'm autistic??'' Hope blossoms within me. \textit{Finally}, I sigh with relief. At last, I will have answers that explain my differences to others.

``\ldots Unfortunately, I'm afraid I don't have the qualifications to diagnose autism. I'd recommend reaching out to a psychiatrist and asking if they have experience with `the female autistic phenotype,' `late-diagnosed autism,' or `autism in adults.'\,'' 

My heart sinks into a pit of despair. Finding a psychiatrist means returning to a mental health system that has repeatedly traumatized me.

\subsection{Narrative 2: Subjugating My Body to the Power of the Medical Industrial Complex\protect\footnote{Mingus describes the ``medical-industrial complex'' as a system ``targeting oppressed communities under the cause of care, health, and safety,'' one ``about profit, first and foremost'' that ``is not just a major piece of the history of ableism, but \textit{all} systems of oppression~\cite{Mingus_2015}.}} \label{narrative-2}

``I'm sorry; I can't help you with that. I work with autistic adults, but I don't perform diagnoses myself.'' 

I stare in exasperation at my new psychiatrist. It's taking everything I have to suppress my internal screams of frustration. ``Well, why not?!''

He gives me a small, sympathetic smile. ``I have neither the training nor experience to do so. Autism research has primarily centered the experiences of white boys, so many practitioners lack the knowledge and skills to diagnose women and other racial groups~\cite{mandell2002race, mandell2009racial, haney2016autism, tsirgiotis2021mixed, travers2018racial}.'' 

Disbelief hits me like a slap to the face. I'm paying \$350 for a psychiatrist to tell me the same thing my therapist has already said? Three weeks spent screening psychiatrists and this was the best I could find? I suppose this is better than the psychiatric nurse who told me, ``There's no way you're autistic. I work with patients who have Autism Spectrum Disorder, and they would never be able to have this conversation that you and I are currently having.'' And this is definitely better than the doctor who pushed sample medications for bipolar disorder on me, despite my insistence that I had never experienced a manic episode.

I decide to take matters into my own hands. ``Look,'' I say. ``Here is the research on how ``masking'' leads to non-stereotypical presentations of autism in adult females~\cite{bargiela2016experiences, Neff_2023, beggiato2017gender, gould2017towards}. I experience the heightened empathy, sensory overload, emotional dysregulation, bottom-up thinking, and hyperfixation described in these studies!''

The pity in my psychiatrist's eyes lets me know that I'm not going to like what he has to say. ``I still think you need a formal evaluation by a neuropsychologist, especially since your preexisting diagnosis of C-PTSD (Complex PTSD) presents additional complications~\cite{dell2023misdiagnoses, carpita2023may}. There's someone who works primarily with refugees, asylum seekers, and immigrants whom I can refer you to. However, she's out-of-network with insurance, so the testing will cost you around \$3,000.''

It's hard for me to hide my anguish and disappointment. ``What's the point of psychiatry then, if you can't help me?'' 

He straightens himself up a little. ``While there is no `cure' for autism~\cite{bolte2014autism, bagatell2010cure}, there are medications I can prescribe to treat your symptoms of anxiety and depression. But something I'd like to discuss is your thoughts on taking ADHD stimulants.''

``What do you mean by that," I respond flatly. 

``Well, watching the way you speak and listening to the stories you shared makes me think that you have ADHD. You are easily overwhelmed by unstructured information; you seem to have trouble sitting still; and you quickly switch between topics, sometimes forgetting what you last said. These traits read as a mixture of executive functioning, hyperactivity, and inattention problems~\cite{wender2001adults, wilens2009presenting}.''

I can't help having optimistic expectations. ``Does this mean that you're diagnosing me with ADHD instead of autism?''

My wishful thoughts are promptly smothered. ``Not exactly,'' he answers. ``You'll have to ask the neuropsychologist about that when you get tested as well. Neuropsychological diagnoses are challenging, because autism and ADHD are co-occurring conditions with a high degree of overlapping symptoms~\cite{gargaro2011autism, stevens2016comorbidity, Neff_2022}. However, the good news is that if your sensory processing issues are related to both ADHD and autism, the stimulant medications can prevent one from exacerbating the other. So what do you say about giving them a try?''

Reluctantly, I agree to trial the ADHD medications, resigning myself to a neuropsychological evaluation. Little did I know that getting a formal diagnosis was not even half the battle.

\subsection{Narrative 3: Accommodating Normativity as a Prerequisite to Accommodating My Neurodivergence} \label{narrative-3}

I take a deep breath, hovering my cursor over the email from my neuropsychologist. My nerves are frazzled, though whether from excitement or trepidation, it's hard to tell. I flick my finger in rapid succession, scrolling down the report until I find the words I'm looking for, ``DSM-V Diagnostic Impressions.''

My heart stops for the brief second that it takes my brain to catch up to my eyes. All at once, waves of satisfaction, relief, and gratitude wash over me. With a bit of pride, I read ``Attention Deficit Hyperactivity Disorder -- Combined Presentation, Autism Spectrum Disorder.''

For the first time in my life, I feel not only seen but validated. Having the words to articulate why I'm different gives me newfound confidence to talk about and normalize my experiences as part of a broader community. I hurriedly reach out to friends and family members to share the good news. 

The response to my diagnosis is generally positive and congratulatory. However, I soon discover that it is much easier to be supportive when all that is required are words and not actions. Over the next several weeks, I am confronted time and again by resistance to my requests for accommodations. 

If I decline attending major celebrations because interacting in large social groups is unbearably painful, I'm ``rude'' and ``hurtful toward their feelings.'' However, if I ask for family events to be hosted near me because traveling is extremely dysregulating, I'm ``inconvenient' and ``ungrateful for their efforts to host me.'' Even the best of intentions can leave me feeling misunderstood and unimportant. 

One weekend, my partner texts me that his friends will be in town tomorrow, and I am invited to join them at the museum and dinner. ``No pressure though,'' he adds. ``It's okay if you're too tired to come; I know big crowds can be overstimulating.''

I accept, despite a rough week at work and issues with my medications, because I \textit{do} enjoy hanging out with his friends, and it \textit{has} been a while since I've seen them.

The following morning, I'm driving back from various errands when I receive a call from my partner. ``Hey, I know this is last minute, but could we meet at the museum an hour earlier? My friends are already in the area.''

Looking at the time, I see that there's barely an hour and a half left until the new meet up time. It'll take me about an hour to get there by public transit, so that leaves me half an hour to get home, eat, shower, and get ready. And that's assuming that I don't take any breaks, forget things, or make mistakes. I pull up the museum website and suggest a compromise, ``How about I join the group midway? I've been to that museum before, so I'm familiar with the main collection. However, I am interested in seeing this temporary exhibit. Could you all save that one for last?''

He agrees, and I'm finishing up my lunch when I receive a text from my partner. ``The museum we originally planned on going to is sold out for the day, so we're heading to this other museum, which is nearby.''

Looking at transit instructions, I see that the bus schedules differ vastly for the two museums, despite their close proximity. My energy reserves are so depleted from juggling the last minute plan changes that I find myself physically incapable of getting up from my couch. I end up skipping the museum to allow my body to rest before dinner. 

I'm waking up from my nap when I receive another text from my partner. ``As a heads up, the restaurant didn't have reservations available, but I spoke to the host, and he said we could probably get walk-in seating fine.''

Looking at restaurant reviews, I see that Saturdays 7 to 9 p.m. are some of their busiest hours. I wince inwardly at the imagined sounds of dishes clashing with each other, servers hollering out orders, music blaring from speakers, and patrons raising their voices to be heard. I also notice that the restaurant serves a cuisine that I tend to avoid because of my hypersensitive nose.

A deep sense of sorrow fills my heart, for this is not the first, nor likely the last time I will remind my partner that to avoid exacting an unbearable toll on my body, I need at least a day's notice for any changes to plans and to be seated in a quiet corner of a sensory-friendly restaurant. It hurts that my neurodivergent needs are continually treated like preferences, often deprioritized in favor of other people's comfort. I want my loved ones to know that when I have to choose between participating in an event that will dysregulate me versus not participating in the event at all, offering me the option to skip the event is not an accommodation. Rather, it sends me the message that ``your needs don't matter enough for us to put time and effort into proactively brainstorming solutions that satisfy all of our needs.'' And although the pain that I feel when my neurodivergent needs are forgotten or ignored can be quite acute, I am unable to express my emotions freely. For the irony is that I cannot receive disability accommodations unless I first communicate my needs in a way that is deemed socially acceptable.

\subsection{Narrative 4: Realizing That My Ability to ``Mask'' and ``Unmask'' My Neurodivergence Is a Privilege} \label{narrative-4}

It's a peaceful Saturday morning, and I am having lunch with my partner when we begin rehashing the same argument that we've had over the past few months. The gist is that I am tired of suppressing my emotions and pretending to be content with the level of compromises my loved ones have offered; I am fatigued by the demands that I must place on my body in order to feel even partially seen or heard; and I am exhausted from adapting to others and forcing myself to fit into the expectations that people have of neurodivergence and disability.  Essentially, I am burnt out from camouflaging my neurodivergent traits through \textit{masking} (``strategies used to hide [neurodivergent] traits or portray a [neurotypical] persona''), \textit{compensation} (``strategies used to actively compensate for difficulties in social situations''), and \textit{assimilation} (``strategies that reflect trying to fit in with others in social situations'')~\cite{hull2019development, belcher2022taking, Neff_2023}. 

I don't know what exactly prompts the sudden change, but all at once, I feel something inside me snap. Tension spews out of my body in the form of hot, angry tears. Before I am able to filter my thoughts, they disperse into the open air in agonizing screams. I have just enough external awareness to grasp that what I'm doing won't make it easier for someone to understand me, but that idea disappears as fast as it appeared. At this point, my body no longer cares how it is perceived; it just seeks relief from the emotional and physical pain. Months of repressed emotions break forth in a fury such that I am unable to comprehend the situation---I am yelling at my partner at the top of my lungs. In between wretched sobs, I shout terrible statements that I would have never said to someone I cared about. For a brief moment, clarity surfaces, and I worry about this horrible fight being how our relationship ends. But the concept of ``future'' has vanished from my mind, and even if it did exist, my concern for consequences is immediately squashed by my desire to be released from present sufferings. My feelings are so intense and overwhelming that verbal communication becomes inaccessible. Words to express my needs are replaced by sounds that defy easy translation; my wails are accompanied by pounding fists and stomping feet.

Later, I learn that I was experiencing a \textit{meltdown}, an ``involuntary response to a nervous system overload''~\cite{Nicole_2023, phung2021wish, lewis2023lived}. As I contemplate the events that transpired, it dawns on me that the behaviors I demonstrated are not as unfamiliar as I originally thought. I have flashbacks to multiple occasions in my childhood when my mother demonstrated similar meltdown behaviors. Additionally, I recall her fixation with routines, sensitivity to touch, and struggle in group settings. I remember the hurt and loneliness that she displayed when others ridiculed and ostracized her for her behaviors. But I also remember the strength and fortitude she possessed to find joy in simple pleasures, such as tending her vegetable garden, while living among people who did not speak her language nor understand her culture. 

As immigrants and refugees, my Vietnamese parents are from a generation that views people with mental health conditions or neurodivergence as ``worthless'' and ``good for nothing''~\cite{vu2014understanding, ha2014living}. Growing up, we never spoke of things such as feelings, and emotional problems were considered a sign of personal failure. It wasn't until my second year in undergraduate study that I braved a venture to the school's psychologist. Although my family experiences contributed to the difficulties I faced in understanding my own neurodivergence, my mother's language and cultural differences actively prevent her from accessing neuropsychological services where she lives. Moreover, these differences affect her ability, or rather inability, to mask her neurodivergent traits. When not speaking English nor understanding Western customs precludes you from socializing with others, blending in is inherently impossible, regardless of neurodivergence. I come to understand that my ability to mask is a privilege, because it grants me access to certain benefits (e.g., employment, safety, and housing) that those who cannot mask face discrimination in~\cite{Juno_2023, Neff_2023Insta, Hammond_2023}.  

However, these benefits don't invalidate the costs that come with masking~\cite{russo2018costs, hull2021social, beck2020looking, bradley2021autistic}. After experiencing a meltdown from excessive masking, I focus on researching steps that I can take to unmask. Developing strategies to live more authentically is both rewarding and exhausting. In the moments that I am able to take off my mask, I feel lighter and freer, but also terrified to be letting go of my safety net. 

While exploring online resources by neurodivergent advocates, I come across the story of Stephon Edward Watts, a 15-year-old Black, autistic teen who loved pomegranates and World of Warcraft, who police killed for holding a butter knife in his own home~\cite{Hurst_2015}. I read about Osaze Osagie, a 29-year-old Black, autistic man who cared deeply for his siblings and his church, who police killed while responding to a mental health-related call~\cite{Rafacz_2020}. I learn about Osime Brown, a 21-year-old Black, autistic man who enjoys eating dumplings and ackee, who faced deportation charges after being falsely convicted of stealing a mobile phone~\cite{Taylor_Busby_2020}. I discover the story about Sheila Jackson's son, a 12-year-old Black, autistic boy who was participating in an after-school tennis program to bond with volunteer officers, when police concussed him for walking away to self-regulate~\cite{Linly_2023}. I hear about Avarius Thompson, a 14-year-old Black, autistic teen who was coming home after buying snacks, when police tased him for being mistaken as a different person~\cite{Chatman_2024}. I find reports about Neli Latson, an 18-year-old Black, autistic teen who  was waiting for the public library to open, when police arrest him for looking like he might have a gun~\cite{Vargas_2010}. Neli becomes triggered and fights back when the officer chokes him, and he receives a ten-year prison sentence. It takes five years before the governor grants Neli a conditional pardon, acknowledging that he needed medical care, not jail time~\cite{Vargas_2020}. But even his full pardon, which only came eleven years after he was sentenced, cannot erase the abuse and trauma that Neli endured in prison~\cite{Arc_2023}. More recently, I read the story of Ryan Gainer, a 15-year-old Black, autistic teen who ran cross-country and dreamt of being an engineer, who police killed for holding a gardening tool~\cite{Levin_2024}. I pore over countless stories by Black and Brown autism advocates who have incredible courage and resilience to continue fighting for intersectional disability justice in a world that does not value their race or ethnicity~\cite{Berne_2015, Biehtar_2023, Hammond_2024, Feral_2022, Beal_2020, Martin_2021, Sam_2020, ObeySumner_2018}. And I come to understand that my ability to unmask is a privilege, because I hold identities that protect me from punishment, violence, and discrimination in access to benefits (e.g., education, social services, and healthcare)~\cite{mccauley2017cumulative, jones2020expert, malone2022scholarly, straiton2022call, jones2020address, tromans2021autism}. 
                                     
\subsection{Narrative 5: Learning to Relate to My Neurodivergence With Self-Compassion} \label{narrative-5}

I awaken to grey, cloudy skies. My body resists the order to fully wake up. Laying in bed, I let the alarm ring until my ears can't stand its incessant shrills anymore. With great reluctance, I drag my covers off my head and ease out of bed. My Apple Watch informs me that ``it's time to start my morning routine!'' I frown in annoyance and dismiss the notification. The Apple Watch was a begrudging purchase that I considered part of my \textit{``ADHD tax,''} the additional financial and emotional costs that I needed to pay in order to do things that are easier for people without ADHD~\cite{Key_2023, sciberras2022social, olsen2022disability}. I had bought it for the sole purpose of helping me establish and maintain routines, but I couldn't find an app that suited my needs. Many apps (e.g., ``Structured - Daily Planner,'' ``Awakee Routine \& Habit Planner'') scheduled routines into a daily calendar, which meant that I had to either follow the exact same routines every day or plan out different routines for specific days ahead of time. Other apps (e.g., ``Fabulous: Daily Habit Tracker'') forced me to adhere to a predefined set of basic routines for weeks before I could focus entirely on my own routines---I immediately uninstalled ``Fabulous'' when I saw that the morning routine first prescribed to me consisted only of drinking water. Some apps (e.g., ``RoutineFlow: ADHD \& Routines'') didn't support watchOS. The app I currently have installed, ``Routinery : Routine \& Planner,'' didn't have any of the above issues, but it still assumed that I would have fixed routines every day and week. However, how I experienced each morning was highly dependent on details that could not be predetermined.

There were some days, like today, when I had less energy, time, or emotional bandwidth. On those days, I learned that I needed to be gentle with myself and accept my body's current state. I would trim my morning routine to only the tasks I needed done before I started my day. There were other days, like yesterday, when I had more energy, time, and emotional bandwidth. On those days, I could challenge myself and incorporate the healthy habits I was practicing into my routine. I would add the ``ideal tasks'' I wanted to complete to my morning routine. My ability to navigate mornings with my neurodivergence existed along a spectrum that was constantly changing. Sometimes the factors that affected my morning routine were entirely unrelated to my body, such as the weather, urgent meetings, or heavy traffic. I wanted a routine app that would be able to accommodate my neurodivergence as a situated experience, adapting to my needs based on context. 

Midmorning, I meet my partner at a local bookstore. As I rush from shelf to shelf, effervescent joy fills my being. It isn't long before I've created a cozy corner for myself, surrounded by stacks of books. When my partner finally catches up to me, my head is buried deep in a novel. ``How are you able to read with your face pressed that closely to the pages?'' he asks with wry amusement.

I glance up in mock horror. ``You're telling me you've never experienced the pure bliss of reading as a multisensory activity? The sight of clean, black lines on crisp, glossy paper? The scent of fresh ink and woodsy pulp on newly printed pages? The smooth, slick texture of each sheet, cooling the skin on a hot summer day??''

His blank expression tells me all I need to know. 

While my sensory processing differences can still be a major constraint in many parts of my life, I have moved away from viewing my neurodivergent traits in absolute terms. In some situations, my neurodivergence is a strength that offers me a unique perspective on life, helping me solve problems in creative ways or savor the moment more deeply. In other situations, my neurodivergence is a limitation that hinders me from my current task, through hurdles such as overstimulation, dysregulation, or executive functioning issues. And although my neurodivergent traits are more likely to be a limitation than a strength, that does not mean that my neurodivergence is inherently negative. Likewise, should my neurodivergent traits ever become more of a strength than a limitation, that does not mean that my neurodivergence is then inherently positive. Rather, I view my neurodivergence as a functional difference that can be adaptive or restrictive, depending on how my body's present conditions interacts with its current surroundings. Learning to accept my neurodivergence without placing value judgments on these differences has made it easier for me to be more understanding and forgiving of myself when difficulties arise.

After our trip to the bookstore, we visit a nearby caf\'e for brunch. I order my favorite dish, the eggs Benedict, and eagerly await my food. Halfway into my fourth bite, I'm struck by an unexpected wave of nausea and forced to set my fork down for the rest of the meal. I bite the inside of my cheek to hold back my tears. Unfortunately, it often happens that finishing a meal can take me over two hours. I've always had a sensitive GI system, and it is exacerbated by my sensory sensitivities and ADHD medication. I resign myself to the fact that I will have to finish the rest of this meal at home. My partner has noticed my expressions and offers me a rueful smile. ``I can provide you with some regulating touch while you eat. Think of it as brunch with a bonus massage.'' 

I've also realized that the instances when my neurodivergent traits are limiting are defined not only by my body state and the environment at that moment but also by normative assumptions about how a ``healthy'' body \textit{should} perform a certain task. For example, taking two or three hours to eat is not disabling if I'm spending the day relaxing at home. But if I'm eating at a restaurant or in back-to-back meetings after the lunch hour, there are social expectations that tables get turned after ninety minutes and meeting attendees not eat on camera. In those cases, my longer eating time becomes disabling, because I'm unable to complete the task in accordance with those norms. Similarly, my inability to communicate verbally during meltdowns is only disabling if someone expects communication to involve words. But that doesn't mean I'm not communicating---\textit{stimming}, short for ``self-stimulatory behavior,'' through repetitive body movements and sounds is a way for me to communicate that I'm currently dysregulated and also a tool to aid me in regulating myself~\cite{kapp2019people, jaswal2019being}. Additionally, stimming is common in all human beings (e.g., fidgeting, singing, humming)~\cite{Lindsmith_2014, Hayden_2022}, and through our relationship, my partner has adopted new forms of self-regulation from watching me stim. \label{stimming} Of course, that's not to say that all social norms are bad and should be eradicated. After all, they do serve a purpose in providing structure and guidance on how we should interact as an interdependent social species~\cite{sunstein1996social, mackie2015social}. Rather, this realization has helped me understand that violating these expectations does not make me ``weird,'' ``inconsiderate,'' ``rude,'' ``annoying,'' ``incompetent,'' ``unworthy,'' or any other label that people have called me. 

My neurodivergence may make me different, but I'm still me, and I'm not letting others define who I am.

\section{Findings} \label{sec:findings}

I now revisit the reflection questions (Section \ref{sec:reflection-questions}), introducing three tenets for reconceptualizing neurodivergence, before discussing the tenets in the context of HCI (Section \ref{sec:discussion}) and future accessibility research (Section \ref{sec:implications}). \ed{Critical autoethnography requires both an intersectional analysis of marginalization and an acknowledgment of privilege through reflexivity~\cite{boylorn2020critical} (Section \ref{sec:critical-autoethnography}). Thus, I analyzed my narratives using the Data Feminism principles articulated by D'Ignazio and Klein~\cite{d2023data} (Appendix \ref{data-feminism-principles}), which are grounded in broader discussions of intersectionality and technology. Moreover, because Data Feminism is an intersectionality framework that operationalizes Critical Theory in a way that is practical, accessible to a wide audience, and inclusive of multiple voices (Section \ref{sec:intersectionality}), my approach aligns with the goal of critical autoethnography "to mobilize and develop the explanatory frameworks that critical theory provides us"~\cite{holman2018creative}.} 

\ed{Each of the following subsections addresses a reflection question in the context of its relevant Data Feminism principles.} In response to RQ1, I suggest that we can reconcile diverse experiences of neurodivergence across different individuals by viewing neurodivergence as a difference, not a deficit (\hyperref[sec:diverse-experiences]{Tenet 1}). With regard to RQ2, I contend that we can affirm individuals' neurodivergent strengths while supporting their limitations by understanding disability as a moment of friction, not a static label (\hyperref[sec:strengths-and-limitations]{Tenet 2}). Finally for RQ3, I propose that we can contribute to a society where all neurodivergent individuals are not only accepted but valued by advocating for accessibility as a collaborative practice, not a one-sided solution (\hyperref[sec:acceptance-value]{Tenet 3}). 

\subsection{Acknowledging Diverse Experiences of Neurodivergence} \label{sec:diverse-experiences}

\ed{The first theme common to my narratives is the diversity of neurodivergent experiences across different individuals. RQ1 seeks to develop recognition for these differences under a conceptualization of neurodivergence that is inclusive of intersectional identities. In my analysis of this theme, I draw on Data Feminism Principles 1, 2, and 5.} As part of ``Principle 1: Examine power''~\cite{d2023data}, D'Ignazio and Klein draw on Patricia Hill Collins' concept of the matrix of domination and \textit{Black Feminist Thought}~\cite{collins1990black} to illustrate ``how systems of power are configured and experienced'' across four domains: the structural (laws and policies), the disciplinary (the implementation of those laws and policies), the hegemonic (culture and media), and the interpersonal (individuals' everyday experiences)~\cite{d2023data}. \label{principle-1} In adherence with this principle, Narratives 1 and 2 highlight how the medical system has the power to decide---via diagnoses---who is considered ``neurodivergent'' (and as a consequence, who is allowed to receive support). Because clinical research has historically focused on white boys~\cite{mandell2002race, haney2016autism, travers2018racial} and efforts to improve participant diversity have primarily considered single-axis frameworks (e.g., either race or gender)~\cite{cascio2021making, mandell2009racial, tsirgiotis2021mixed}, my intersectional experiences of neurodivergence were marginalized by medical professionals. At the same time, Narratives 1 and 2 also surface some of the privileges I held to be able to fight against this system. For example, my employment status and benefits gave me the resources (e.g., time and money) to visit multiple professionals. I did not have to deal with the cultural and language barriers that neurodivergent immigrants face in accessing health care~\cite{lim2021review, rosenberg2020disparities}. And as a cisgender woman, I did not have to worry about medical providers misusing my neurodivergence as a reason to deny gender-affirming care~\cite{maroney2022tuned}.

In Narrative 4, I expand upon this analysis of power, incorporating ``Principle 2: Challenge power``~\cite{d2023data}. D'Ignazio and Klein present four starting points for taking action: (1) \textit{Collect}: ``compiling counterdata---in the face of missing data or institutional neglect,'' (2) \textit{Analyze}: ``demonstrating inequitable outcomes across groups,'' (3) \textit{Imagine}: ``imagin[ing] our end point not as `fairness,' but as co-liberation,'' and (4) \textit{Teach}: engag[ing] and empower[ing] newcomers to the field in order to shift the demographics and cultivate the next generation''~\cite{d2023data}. \label{principle-2} Stephon Edward Watts's, Osaze Osagie's, Osime Brown's, Sheila Jackson's, Avarius Thompson's, Neli Latson's, and Ryan Gainer's stories demonstrate how multiple social institutions work together to render intersectional, neurodivergent experiences invisible. Diagnostic referrals tend to come from the education system, which treats behavioral differences in white children as medical concerns while punishing Black and Brown kids for the same behaviors~\cite{tromans2021autism, skiba2011race, macsuga2021disproportionate}---a disparity that continues as students age, and Black and Brown neurodivergent individuals become disproportionate targets of police violence~\cite{hutson2022m} and institutionalized torture~\cite{neumeier2019torture} instead of social support.
  
Narrative 4 also engages with ``Principle 5: Embrace pluralism,'' by ``synthesizing multiple perspectives, with priority given to local, Indigenous, and experiential ways of knowing''~\cite{d2023data}. \label{principle-5} Without those additional stories, my racial privilege and the systemic oppression of neurodivergent, Black women in clinical research~\cite{malone2022scholarly, radulski2022conceptualising, diemer2022autism, lovelace2021missing} would have led to a more reductionist discussion about the harms of masking, one that does not acknowledge the privileges that come with the ability to mask and unmask~\cite{Hammond_2023}. As part of Principle 5, D'Ignazio and Klein explain that ``all knowledge is partial, meaning no single person or group can claim an objective view of the capital-T Truth''~\cite{haraway2016situated, d2023data}. They recommend practicing transparency and reflexivity by disclosing one's own \textit{positionalities}, or ``intersecting aspects of any particular person's identity''~\cite{d2023data}. Although critical autoethnography as a method already requires reflexivity (Section \ref{sec:critical-autoethnography}), autoethnography's evocative nature inherently centers the narratives around the author's limited perspective~\cite{lapadat2017ethics}. Therefore, in addition to disclosing my positionalities (Section \ref{sec:limitations}), I encourage readers to engage directly with the first-person perspectives that influenced this work, including Indigenous~\cite{Ashburn_Edwards_Onaiwu_McConnell_2023, Biehtar_2023, Brown_Ashkenazy_Onaiwu_2017}, Black~\cite{Brown_Ashkenazy_Onaiwu_2017, Hammond_2023}, queer~\cite{Brown_Ashkenazy_Onaiwu_2017, clare2017brilliant}, and nonspeaking~\cite{NeuroClastic_2021, Tziavaras_2021, Whitty_2021} neurodivergent voices. \label{underheard-voices}

When I started my neurodivergent journey, I was fixated on proving my neurodivergence (\hyperref[narrative-2]{Narrative 2}), clinging to my diagnoses of autism and ADHD as evidence that I wasn't ``broken'' (\hyperref[narrative-3]{Narrative 3}). However, as I learned about the wider neurodivergent community, I realized that giving diagnostic criteria that much weight and importance meant that I was erasing the experiences of other intersectional, neurodivergent individuals (\hyperref[narrative-4]{Narrative 4}). To shift power away from diagnoses and their deficit-based criteria, I offer \textbf{Tenet 1: Neurodivergence is a functional difference, a set of unique traits that can be adaptive depending on context}.

\subsection{Supporting Both Neurodivergent Strengths and Limitations} \label{sec:strengths-and-limitations}
\ed{The second theme common to my narratives is the existence of both neurodivergent strengths and limitations. RQ2 aims to understand how we can honor the nuances in these experiences through a more holistic and compassionate conceptualization of neurodivergence. Data Feminism Principles 3 and 4 guide my exploration of this theme.} Thus far, the examination of power in Narratives 1, 2, and 4 has focused on the structural and disciplinary domains. In Narrative 3, I illustrate how ideas about ``socially-acceptable behavior'' (hegemonic domain) led to experiences of rejection and abandonment by my loved ones (interpersonal domain). Social norms are an example of universal rules that ``exclude some aspects of human experience in favor of others''~\cite{d2023data}. To counteract these harms, D'Ignazio and Klein offer ``Principle 3: Elevate emotion and embodiment,'' which ``teaches us to value multiple forms of knowledge, including the knowledge that comes from people as living, feeling bodies in the world'' ~\cite{d2023data}. \label{principle-3} I engage with this principle in Narrative 3, when I reflected on the lack of empathy I continued to receive after my diagnosis. This prompted me to question behavioral definitions and norms, especially when I was often asked to understand why someone couldn't accommodate me, despite supposedly lacking empathy as an autistic person~\cite{milton2022double}. I wondered why nonspeaking neurodivergent individuals or those with co-occurring intellectual disabilities were deemed less intelligent or incapable of communication, unless they proved their abilities in a way society could understand~\cite{Baggs_2007, Ashburn_Edwards_Onaiwu_McConnell_2023, clare2017brilliant}. Does empathy truly mean to ``the ability to understand another person's feelings, experience, etc.''\footnote{Definition obtained from the Oxford Advanced American Dictionary}, if we also require people to behave in ways we can comprehend (e.g., communicate with words) before we are willing to empathize with them? 

In ``Principle 4: Rethink binaries and hierarchies,'' D'Ignazio and Klein explain how counting and classification ``have been used [as tools of power] to dominate, discipline, and exclude''~\cite{d2023data}. Narrative 5 challenges a binary understanding of neurodivergence, by rejecting the notion that certain neurodivergent traits are positive whereas others are negative. Using my sensory processing differences as an example, I illustrate how my neurodivergence can be both a strength (\hyperref[narrative-5]{Narrative 5}) and a limitation (\hyperref[narrative-1]{Narrative 1}). In resisting deficit-based perceptions of neurodivergence, I am also challenging the disabled/non-disabled binary that medicalizes disability as something to be ``cured.'' Through Narrative 5, I demonstrate how ``disabled'' is a state that changes over time, based on both internal and external factors (e.g., energy level, location, and work demands). Likewise, neurodivergence does not always imply disability and can even be adaptive in some situations, such as the use of stimming for self-regulation. To avoid portraying intersectional, neurodivergent experiences as ``problems to be solved,'' Narrative 4 introduces each Black or Brown neurodivergent individual as a person with ordinary wishes and dreams, before discussing how they were subjugated to police brutality or prisoner abuse. Even then, Black and Brown neurodivergent communities are not passive victims in need of ``saving''; they have been using their strengths and agency to fight against intersectional injustice~\cite{Berne_2015, Biehtar_2023, Hammond_2024, Feral_2022, Beal_2020, Martin_2021, Sam_2020, ObeySumner_2018}.

Throughout my neurodivergent journey, I struggled to convey a more holistic understanding of my neurodivergence, one that captured the nuanced aspects of my identity. Influenced by society's neuronormative attitudes, others treated my neurodivergent traits as burdens to overcome, rather than constraints to inspire creativity (\hyperref[narrative-3]{Narrative 3}). I realized that my neurodivergence was not inherently disabling; the situations in which I thrived depended both on my current capacities and the existence of a supportive environment, where people were willing to accommodate my neurodivergence---including temporal fluctuations and complex manifestations---and explore alternative approaches to an activity (\hyperref[narrative-5]{Narrative 5}). This leads me to present \textbf{Tenet 2: Neurodivergent disability is a situated, embodied, and dynamic experience, mediated by the interaction between an individual's current abilities and their present environment}.

\subsection{Moving Beyond Acceptance to Recognize the Value of Neurodivergence} \label{sec:acceptance-value}
\ed{The last theme common to my narratives is the desire for neurodivergence to not only be accepted but valued by society. RQ3 contends with the challenges that social systems present regarding this desire. To investigate this theme, I leverage Data Feminism Principles 6 and 7.} In ``Principle 6: Consider context,'' D'Ignazio and Klein ``assert that data are not neutral or objective. They are the products of unequal social relations, and this context is essential for conducting accurate, ethical analysis''~\cite{d2023data}. Engaging with this principle means recognizing that deficit-based perceptions of neurodivergence are a product of broader social systems. Black performance artist Tricia Hersey argues in \textit{Rest is Resistance} that grind culture and ``what we have internalized as productivity has been informed by a capitalist, ableist, patriarchal system'' that ``leads us to the path of exhaustion, guilt, and shame''~\cite{hersey2022rest}. Indeed, it is this drive to keep up with school and work demands that leads many neurodivergent women, including myself, to mask our neurodivergence to the point of burnout~\cite{Honeyman_2024, belcher2022taking}. Narrative 1 and 2 illustrate how failing to consider context led me to view my neurodivergent traits as inherent limitations. Because society treats productivity as a virtue, I pursued my late-diagnoses of ADHD and autism with medicalized notions of my body as ``problematic'' and ``defective'' (\hyperref[narrative-1]{Narrative 1}). Internalized ableism\footnote{Shew explains ableism as ``bias against disabled people, bias in favor of nondisabled ways of life''~\cite{shew2023against}.} also showed up in the way I relied heavily on medication and diagnostic tests to validate my socioemotional and executive functioning differences as autistic and ADHD symptoms (i.e., deficits)~\cite{kattari2018you}, rather than natural variations as part of neurodiversity~\cite{kapp2013deficit} (\hyperref[narrative-2]{Narrative 2}).

When I treated my neurodivergence as a problem, I not only underestimated the value of my neurodivergent traits; I also underestimated my own self-worth. Narrative 3 highlights how others' normative expectations take my energy and effort for granted, demonstrating the importance of ``Principle 7: Make labor visible''~\cite{d2023data}. In this last principle, D'Ignazio and Klein insist that ``undervalued and invisible labor receives the credit it deserves''~\cite{d2023data}. In addition to the invisible labor required to even attend social gatherings, Narrative 3 also draws attention to my \textit{emotional labor}, ``the work involved in managing one's feelings, or someone else's, in response to the demands of society or a particular job''~\cite{d2023data}. Regardless of how exhausted I felt inside, I carefully monitored my affect to ensure I seemed interested and engaged, lest I hurt another person's feelings. I also had to manage the painful emotions I felt when others assumed that I didn't want to or couldn't participate in an activity just because I had never taken part previously. Yes, I tend to avoid social gatherings, but that's not because I don't like people. It's because they typically involve loud noises and hours of small talk that my body can't endure. When people make these limiting assumptions, they act in ways that reinforce these beliefs (e.g., not inviting me to group events, inviting me but leaving me to read in a corner).

Even when people accepted my neurodivergent traits as differences versus deficits and collaborated with me to reduce moments of disability, my neurodivergence was still regarded as something to be supported, rather than something that could be supportive (\hyperref[narrative-3]{Narrative 3}). Indeed, it wasn't until my partner let go of neuronormative expectations that things changed (\hyperref[narrative-5]{Narrative 5}). When my partner stopped trying to assist me with social norms and started adapting his behaviors to interact in ways that were more natural to me, we discovered new ways of connecting with each other that defied normative ideas about language and physical touch yet allowed us to meet our needs with greater ease. Of course, it isn't possible to change society overnight, so we couldn't disregard social norms in all interactions, but in the situations where it was possible, I felt not only that I was included in the interaction but also that I belonged. As a step toward a more inclusive society, I offer \textbf{Tenet 3: Neurodivergence accessibility involves embracing neurodivergent differences as valuable and learning from them to generate transformative solutions}. 

\section{Discussion} \label{sec:discussion}
The previous section analyzes the reflection questions (Section \ref{sec:reflection-questions}) in relation to the autoethnographic narratives, deriving three tenets for reconceptualizing neurodivergence via Data Feminism as the intersectional lens. In this section, I apply the same intersectional framework to an analysis of the reflection questions in HCI (using an identical mapping between tenet and reflection question as described in Section \ref{sec:reflection-questions}), demonstrating the applicability of the tenets to research. In the next section, I provide specific implications for engaging the tenets in future accessibility work (Section \ref{sec:implications}).

\subsection{Tenet 1: Neurodivergence as a Difference, Not a Deficit} \label{sec:tenet-1}
\textbf{Neurodivergence is a functional difference, a set of unique traits that can be adaptive depending on context.}

As shown in Section \ref{sec:diverse-experiences}, neurodivergent diagnoses are gated by systems that marginalize intersectional, neurodivergent identities. Power is also distributed across multiple, interlocking systems (e.g., medical, education, and government), which is why efforts to improve the gender and racial diversity in clinical diagnostic research continue to exclude Black, queer, and trans women, with co-occurring conditions and who live in low-income households~\cite{lovelace2021missing, cascio2021making, Brown_Ashkenazy_Onaiwu_2017}. With regard to HCI, applying ``Principle 1: Examine power'' involves reflecting on questions such as ``Who does the work (and who is pushed out)? Who benefits (and who is neglected or harmed)? Whose priorities get turned into products (and whose are overlooked)?''~\cite{d2023data}. In line with the findings above, prior work has demonstrated that HCI neurodivergence research has mainly focused on the experiences of children~\cite{spiel2019agency} and ``male, English speaking adults between 20 and 30 years old living in the USA and Europe''~\cite{ramirez2024adult}. \ed{Clinical diagnostic criteria's continued influence in medicine and HCI creates challenges for intersectional, neurodivergent identities to be seen---in my case, this showed up as difficulties with obtaining a formal diagnosis (Narratives \hyperref[narrative-1]{1} \& \hyperref[narrative-2]{2}) and lack of validation from my loved ones (\hyperref[narrative-3]{Narrative 3}).}

In terms of ``Principle 2: Challenge power''~\cite{d2023data}, disabled and neurodivergent scholars have already been taking action by collecting and analyzing data on trends in overall accessibility research~\cite{mack2021we} as well as HCI neurodivergence research, demonstrating the dominance and harms of behavioral intervention technologies (e.g., ~\cite{williams2020perseverations, spiel2019agency, spiel2021purpose, spiel2022adhd, hundt2024love, keyes2020automating, ramirez2024adult}). A further examination of power (\hyperref[principle-1]{Principle 1}) reveals that ``the multi-billion-dollar behaviorist industry has a financial interest in pathologizing autistic traits in order to sell their services of producing non-defective children''~\cite{Ashburn_Edwards_Onaiwu_McConnell_2023}, which is why ``ABA [is] the `only' `evidence-based' and therefore only health-insurance-fundable, intervention for autism''~\cite{broderick2021autism}. Pediatric psychologist Mona Delahooke elaborates, ``Too often, professionals label a child's behavioral differences as part of a checklist of the autism diagnosis rather than seeing them as adaptations to how information is processed through the child's body/brain information highway''~\cite{delahooke2020beyond}. Delahooke's explanation supports neurodivergent individuals' assertion that behavior is communication~\cite{Ashburn_Edwards_Onaiwu_McConnell_2023, Baggs_2006}. \ed{HCI's historical focus on social skills interventions for autistic children~\cite{spiel2019agency} mirrors the healthcare industry's emphasis on ABA therapy as well as societal expectations for neurodivergent individuals to engage in masking, compensation, and assimilation. These practices can have detrimental effects, including increased stress~\cite{zolyomi2019managing}, traumatic memories~\cite{anderson2023autistic}, and meltdowns~\cite{phung2021wish}---such as the one described in \hyperref[narrative-4]{Narrative 4}.}

Based on the concept of neurodiversity~\cite{walker2021neurodiversity}, Tenet 1 provides researchers with an alternative to diagnostic labels that is more inclusive of diverse experiences of neurodivergence across intersectional identities. By reconceptualizing neurodivergence as a functional difference, researchers can remain open to how neurodivergence manifests as varying mental and physical capabilities based on context\ed{, which is evidenced by how my body state and emotional bandwidth fluctuates depending on the morning (\hyperref[narrative-5]{Narrative 5}).} Viewing neurodivergence as a difference versus a deficit also aligns with the lived experiences of other neurodivergent individuals~\cite{kapp2013deficit}. Additionally, without the limitations of diagnostic criteria, researchers can engage with ``Principle 5: Embrace pluralism'' and ``remember and acknowledge multiple, even contradictory versions of reality''~\cite{d2023data}, which exist in the neurodivergent community (e.g., between autistic self-advocates and autism parents~\cite{Ashburn_Edwards_Onaiwu_McConnell_2023}, BIPOC versus white autistic individuals~\cite{Brown_Ashkenazy_Onaiwu_2017}). Whereas diagnoses have been used as a ``weapon to shut down conversations [and] bully others''~\cite{Hammond_2024d}, Tenet 1 supports multiple perspectives on neurodivergence, which autism mom Ashburn and autistic parent Edwards have demonstrated promotes ``conflict in a way that builds relationships and invites others to join us in our work''~\cite{Ashburn_Edwards_Onaiwu_McConnell_2023}. 

\subsection{Tenet 2: Neurodivergent Disability as a Moment of Friction, Not a Static Label} \label{sec:tenet-2}
\textbf{Neurodivergent disability is a situated, embodied, and dynamic experience, mediated by the interaction between an individual's current abilities and their present environment.}

Section \ref{sec:strengths-and-limitations} highlights how normative assumptions treat neurodivergence and disability as ``fitting into neat arrangements,'' disregarding the ways individuals are shaped by their surroundings~\cite{whittaker2019disability}. Social norms also perpetuate the false notion that disabled people are inferior to non-disabled people~\cite{Jawadi_2022}, despite the fact that disability is an fundamental part of being human~\cite{WHO_2020, shew2023against}. Nonspeaking neurodivergent individuals have explained how neuronormative beliefs are harmful, as they result in ``support [that] is often provided in ways that strip away a person's autonomy and humanity\ldots based on the assumption that the person being supported is fundamentally incapable of handling a certain task''~\cite{Ashburn_Edwards_Onaiwu_McConnell_2023, stout2020presuming, NeuroClastic_2021}. Instead, they advocate for \textit{presumed competence}, because an ``assumption of intelligence brings respect, whether that intelligence is on display or not, \ldots and respect brings dignity''~\cite{Sarathy_2019, stout2020presuming, Ashburn_Edwards_Onaiwu_McConnell_2023}. \ed{Similar to how my loved ones assumed that I have no interest in social events (\hyperref[narrative-3]{Narrative 3}), HCI researchers have predominantly designed games for neurodivergent people that are single-player with a medical or educational purpose, based on an assumption of neurodivergent asociality~\cite{spiel2021purpose}. These beliefs are harmful, because they fail to recognize how neurodivergent individuals also desire to experience play, enjoyment, and connectedness---in this way, we are no different from other human beings.}

Applying ``Principle 3: Elevate emotion and embodiment'' to HCI entails acknowledging that science has often used objectivity, rationality, and neutrality as tools to suppress marginalized voices~\cite{d2023data, haraway2016situated}. In the context of neurodivergence, researchers have rejected the idea of presumed competence for nonspeaking individuals or those with co-occurring intellectual disabilities, arguing that the premise is grounded in emotion rather than evidence-based practices~\cite{travers2015critique, o2018pitfalls}. However, as demonstrated in Section \ref{sec:tenet-1}, such critiques fail to acknowledge the ways ``evidence-based practices'' are rooted in structural oppression and ``often synonymous with white-normed research''~\cite{manns2023exploring}. As an alternative to presumed competence, Travers and Ayres suggest ability agnosticism, which relies on the assumption that people are able to judge someone's ability without prejudice or bias~\cite{travers2015critique}. However, Williams et al. have shown that without critical engagement with perspectives from disabled and neurodivergent communities, researchers' judgments of ability will be inherently biased against non-normative behaviors~\cite{williams2023counterventions}. \ed{Indeed, Spiel et al. demonstrated that the mere inclusion of ADHD people in user-centered design, participatory design, or co-design approaches can still lead to detrimental outcomes, especially when researchers continue to privilege the neurotypical perspectives of parents, educators, or medical professionals~\cite{spiel2022adhd}. Recognizing the embodied nature of neurodivergent disability means empowering neurodivergent individuals with the agency to define the situations in which they are disabled and the factors that disable them during those moments. These factors can be internal, external, or a mixture of both: my gastrointestinal sensitivities become disabling when I am expected to finish a meal before my next meeting or within a restaurant's specific time frame (\hyperref[narrative-5]{Narrative 5}).}

Coinciding with enactivism and other interactional approaches to HCI~\cite{bennett2021emergent, boehner2005affect}, Tenet 2 frames neurodivergent disability as an emergent situation that arises from the complex interplay between an individual and their surroundings. In doing so, Tenet 2 incorporates ``Principle 4: Rethink binaries and hierarchies''~\cite{d2023data}, challenging the disabled/non-disabled binary by reconceptualizing disability as a ``complex web of interactions, relations, and identities''~\cite{howard2017beyond}. Tenet 2 enables researchers to attend to the environmental and task factors that shape perceptions of neurodivergent traits as strengths or limitations, supporting the neuro-positionality framework's recommendation to ``champion dynamic adaptability''~\cite{stark2023disabling} as well as prior HCI work demonstrating that access is varied, contextual, and overlapping across multiple domains~\cite{mack2022chronically, spiel2022adhd, williams2023cyborg, borsotti2024neurodiversity}. \ed{This is particularly relevant for the inclusion of people with intersectional, neurodivergent identities, such as my mother and the Black and Brown individuals mentioned in \hyperref[narrative-4]{Narrative 4}, whose race or ethnicity resulted in additional cultural and language barriers.} Moreover, by aligning with Shew's argument that ``\,`no trait can be thought of as being good or bad independently' of environment, response, and task''~\cite{shew2023against, macintyre1979seven}, Tenet 2 is an assertion of the common humanity that neurodivergent people share with neurotypical people---what shame researcher Bren\'e Brown describes as our shared struggle to ``let go of who we think we're supposed to be and fully embrace our authentic selves---the imperfect, creative, vulnerable, powerful, broken, and beautiful''~\cite{brown2022gifts}. 

\subsection{Tenet 3: Neurodivergence Accessibility as a Collaborative Practice, Not a One-Sided Solution}
\label{sec:tenet-3}
\textbf{Neurodivergence accessibility involves embracing neurodivergent differences as valuable and learning from them to generate transformative solutions.} 

Section \ref{sec:acceptance-value} draws attention to how negative attitudes about neurodivergence are embedded in racism, classism, capitalism, and other systems of oppression, which undervalues neurodivergent labor. In the context of HCI, engaging with ``Principle 6'' means recognizing that societal values defining what is ``good'' or ``bad'' do not exist in the abstract~\cite{menakem2021my}. Similar to perceptions of neurodivergence, the values that drive technology research are also not neutral. Lin and Lindtner explain how HCI's ``value system of usefulness (of technology as enabler of social and human progress, productivity, and excellence)'' puts the focus on ``\textit{how to use something} but not \textit{who can use something},'' perpetuating normative and harmful ideas about able bodies that mask structural issues as personal problems~\cite{lin2021techniques}. \ed{This is similar to how I viewed my autistic burnout as a problem with myself rather than capitalism, leading me to pursue a medical diagnosis rather than question my stressful work environment (\hyperref[narrative-1]{Narrative 1}).} In their ``counterventions'' framework~\cite{williams2023counterventions}, Williams et al. demonstrate how an analysis of context is crucial to avoiding \textit{technoableism}, termed by Shew as ``the harmful belief that technology is a `solution' for disability''~\cite{shew2023against}. \ed{How we frame problems affects where we look for solutions—regarding neurodivergence, this has led to curative practices in medicine and HCI.}

Tenet 3 rejects the harmful assumption that neurodivergent individuals are only passive recipients of disability accommodations or worse, a burden on social resources~\cite{Ashburn_Edwards_Onaiwu_McConnell_2023, Brown_Ashkenazy_Onaiwu_2017, robb2011estimated}. Building upon interdependence for assistive technology design~\cite{bennett2018interdependence}, Tenet 3 asserts that neurodivergence accessibility is predicated on mutually beneficial collaborations that value neurodivergent traits and resist neuronormative expectations. Learning from non-normative ways of interacting can help researchers challenge power and imagine neurodivergence accessibility as co-liberation (\hyperref[principle-2]{Principle 2}). Valuing neurodivergence also addresses ``Principle 7: Make labor visible''~\cite{d2023data}, as it includes honoring the labor that neurodivergent people perform. As HCI shifts toward more participatory and community-based research, it is important that researchers recognize the kinds of unpaid labor that neurodivergent and disabled individuals perform in these engagements. Recognizing that conversations around intersectionality and disability often happen outside of academia, Harrington et al. recommend that researchers learn from these activist-led projects~\cite{harrington2019deconstructing}. At the same time, Guberman draws attention to how researchers have used autistic social media data without authors' consent or respect, `` \,`treating [their] writing like a zoo exhibit\,' ''~\cite{guberman2023actuallyautistic}. Likewise, Hayes highlights how community partners are left out of conferences and authorship recognition on publications they contributed to and ``exploited, like natural resources for the harvesting''~\cite{Hayes_2019}. In honoring the value of neurodivergent knowledge and expertise, Tenet 3 encourages HCI researchers to heed Cooper et al.'s suggestions to``partner with participants to share the results of the research'' and ``monitor and maintain community relationships''~\cite{cooper2022systematic}. \ed{Sustainable partnerships with neurodivergent populations requires recognizing the invisible labor that neurodivergent individuals put into relational interactions---in my narratives, I describe how my partner moved from simply acknowledging my neurodivergence (\hyperref[narrative-3]{Narrative 3}) to actively collaborating with me during moments of disability (\hyperref[narrative-5]{Narrative 5}).}

\section{Implications for Research} \label{sec:implications}
Based on the discussion of the tenets in relation to HCI (Section \ref{sec:discussion}), I offer the following suggestions for addressing the reflection questions (Section \ref{sec:reflection-questions}) in future research. Namely, researchers could acknowledge diverse experiences of neurodivergence (RQ1) by using Tenet 1 to \ed{investigate when neurodivergent traits are adaptive and how that differs across individuals from diverse backgrounds} (Section \ref{sec:imp-1}). To affirm neurodivergent individuals' strengths while supporting their limitations (RQ2), researchers could use Tenet 2 and \ed{actively challenge behavioral norms when building technology} (Section \ref{sec:imp-2}). Lastly, researchers could contribute to a society where neurodivergent individuals are not only accepted but valued (RQ3) by using Tenet 3 to design technology for mutual benefit and co-liberation (Section \ref{sec:imp-3}).

\subsection{Investigate the Diverse Contexts in Which Neurodivergent Traits are Adaptive} \label{sec:imp-1}
\textbf{Tenet 1: Neurodivergence is a functional difference, a set of unique traits that can be adaptive depending on context.}

Neurodivergent scholars have urged HCI to move away from the deficit-based perceptions of neurodivergence that have dominated technology research~\cite{spiel2019agency, spiel2021purpose, spiel2022adhd, williams2020perseverations, williams2023counterventions}. \ed{To address these calls to action, researchers could use Tenet 1 to reconceptualize neurodivergent traits as adaptive differences versus diagnostic symptoms. In my narratives, I demonstrate how I have moved from judging myself harshly based on diagnostic criteria (\hyperref[narrative-1]{Narrative 1}) to being open-minded and curious about how my neurodivergent traits allow me to uniquely navigate the world, such as the use of stimming as a form of self-regulation and communication (\hyperref[narrative-5]{Narrative 5}). This creative outlook has allowed me to accept and value my neurodivergence, by letting go of the need to fit into society's neuronormative expectations.}

\ed{Although HCI researchers are shifting from deficits- to strengths-based approaches (e.g.,~\cite{boyd2023conceptualizing, williams2023cyborg, hall2024designing}), these works discuss strengths at a more abstract level (e.g., values~\cite{hall2024designing}, domains~\cite{williams2023cyborg}) than functional traits (e.g., focus levels, communication styles). Moreover, while Boyd's paradigm of ``Celebratory Technology for Neurodiversity'' also argues for ``neurodiversity as a natural difference and not less than neurotypical''~\cite{boyd2024designing}, Tenet 1 adds nuance to this idea by suggesting that the \textit{adaptivity of neurodivergent traits is context-dependent}---my narratives show that masking and sensory sensitivities are not always strengths (Narratives \hyperref[narrative-1]{1} \& \hyperref[narrative-4]{4})---a distinction that could mitigate the ``risk of celebratory efforts being perceived as `inspiration porn'\,''~\cite{boyd2023conceptualizing}. Applying Tenet 1, researchers might consider investigating when neurodivergent traits are adaptive and how that differs across individuals from diverse backgrounds.} 

As part of including intersectional experiences of neurodivergence, researchers might also consider recruiting participants who are \textit{``neurodivergent questioning,''} a term I adapt from gender questioning to refer to ``person who is in the process of figuring out how to describe and label their [neurocognitive functioning], but has reason to think that they might be [neurodivergent]''~\cite{Nonbinary_Wiki_2023}. This practice would build upon recent work in HCI studying self-diagnosis and neurodivergence~\cite{alper2023tiktok, eagle2023you}---and counteract the biases inherent in diagnostic criteria, which as Section \ref{sec:tenet-1} illustrates, have led to the exclusion gender, racial, and other minority group identities from HCI neurodivergence research~\cite{ramirez2024adult}. In addition to reducing the influence of diagnostic criteria, accepting self-diagnosed and questioning neurodivergent individuals is a form of valuing lived experience and affirming neurodivergent self-expertise~\cite{sarrett2016biocertification} (\hyperref[principle-3]{Principle 3}). Labels such as neurodivergence and disability are constantly evolving in response to contemporary scholarship~\cite{sarrett2016biocertification}, so it's important that researchers stop treating them as ``essential, fixed classifications''~\cite{whittaker2019disability}.

\subsection{Actively Challenge Behavioral Norms When Building Technology} \label{sec:imp-2}
\textbf{Tenet 2: Neurodivergent disability is a situated, embodied, and dynamic experience, mediated by the interaction between an individual's current abilities and their present environment.}

\ed{While Critical Disability Studies has developed various conceptual models to explain disability in ways that transcend the social model/medical model binary (e.g.,~\cite{botha2021critical, davidson2010cuts, sarrett2012autistic, kapp2013interactions, anderson2022autism, kintzinger2021equity, chapman2021neurodiversity, maiese2021enactivist, nevsic2023ecological}), these models of disability have been criticized for failing to describe how they can be operationalized practically~\cite{shakespeare2013disability}. Building on prior work to bridge the gap between accessibility theory and practice~\cite{mcdonnell2023tackling}, Tenet 2 uses Tenet 1's understanding of neurodivergence as a functional difference, not deficit, to reconcile the perceptual misalignment between dominant perspectives on neurodivergence (i.e., the seeming contradictions between the social and medical models of disability). Neurodivergent disability is not a static problem located in either the individual or their environment. Instead, disability arises from the situated, embodied, and dynamic interaction between a specific individual's abilities and their current environment. Neurodivergent disability is a \textit{moment of friction separate from the individual and their environment}. Thus, disability presents a collaborative problem-solving opportunity for both the disabled individual and their environment.} 

\ed{Seen this way, the social and medical models of disability are separate but not mutually exclusive tools for navigating disability---one targets changes to the environment whereas the other targets changes to the individual, respectively. In \hyperref[narrative-5]{Narrative 5}, I highlight how I've applied Tenet 2 in my own life: Rather than stick to ADHD apps that required fixed routines every day and week, I adjusted my morning routine each day, adapting to the internal (e.g., energy, emotional bandwidth) and external (e.g., weather, traffic) factors that manifested my experiences of disability. Applying Tenet 2 to accessibility research, researchers might consider building technological supports that accommodate contextual changes in both the individual and their environment. Researchers could also adapt Tenet 2's understanding of disability to other disability communities beyond neurodivergence.}

\ed{Additionally, my narratives show that neurodivergent disability can arise from perceiving situations through normative expectations. In the example of sharing a meal with my partner, I experienced neurodivergent disability when he assumed that I could tolerate last-minute plans and sensory overload (\hyperref[narrative-3]{Narrative 3}). However, despite having gastrointestinal sensitivities that made eating take two to three hours, I did not experience neurodivergent disability when he offered me regulating touch and a change in location (\hyperref[narrative-5]{Narrative 5}).} Likewise, Section \ref{sec:tenet-2} highlights how presumed competence contributes to an understanding of neurodivergent disability that affirms neurodivergent individuals' strengths while supporting their limitations. \ed{Therefore, another application of Tenet 2 could involve actively challenging behavioral norms when designing technology.} 

\ed{Although HCI has increasingly accepted neurodivergent behaviors (e.g.~\cite{zolyomi2023designing, morris2024understanding, zolyomi2024emotion}), a neurodivergent/neurotypical binary is reinforced through concepts such as ``emotion translator''~\cite{zolyomi2024emotion} or the double empathy problem~\cite{morris2024understanding}, which portray neurodivergent and neurotypical communication as fundamentally different. Instead, researchers could investigate how technology can create sensory-friendly/immersive environments (e.g., via extended reality (XR)) where mixed neurotype individuals can co-create new forms of sense-making beyond words or facial expressions---similar to how my partner and I used stimming to connect in my narratives (\hyperref[narrative-5]{Narrative 5}).} By resisting ideas that ``other'' neurodivergent individuals, Tenet 2 mitigates the risks of empathy in human-centered design, which reinforce a binary between ``designing bodies as non-disabled bodies \ldots [and] disabled bodies as non-designing bodies''~\cite{bennett2019promise}. When people have acknowledged the inherent value in neurodivergent and disabled individuals' natural behaviors while recognizing the challenges they face, neurodivergent and disabled individuals have proved society's low expectations wrong over and over again~\cite{Ashburn_Edwards_Onaiwu_McConnell_2023, Sarathy_2019, Johnson_2015, Tziavaras_2021, Hammond_2024c, Whitty_2021, Brown_Ashkenazy_Onaiwu_2017}.

\subsection{Design for Co-Liberation (Mutual Benefit)} \label{sec:imp-3}
\textbf{Tenet 3: Neurodivergence accessibility involves embracing neurodivergent differences as valuable and learning from them to generate transformative solutions.} 

Prior work has highlighted the importance of lived experiences in accessibility research (e.g., ~\cite{mankoff2010disability, spiel2020nothing, mack2022anticipate, williams2023counterventions, mack2021we}), challenging ``the assumption that autistic people are incapable of contributing expert user knowledge to the design process of new technologies''~\cite{elmore2014embracing}. As referenced in Section \ref{sec:tenet-3}, Bennett et al.'s interdependence frame ``establishes [disabled people] as contributors to---not just recipients of---community support and assistance''~\cite{bennett2018interdependence}. However, interdependence in HCI neurodivergence research has primarily examined how more cooperative practices can benefit neurodivergent individuals in mixed neurotype\footnote{\textit{Neurotype} refers to ``a certain way the brain works'' in terms of ``neuro-cognitive variability or styles''~\cite{Boren_2024}.} interactions~\cite{borsotti2024neurodiversity, boyd2024neuro, liu2023survey, bayor2021toward, sobel2016incloodle}, contributing to the trend where ``assistive technology interventions target an individual already overly burdened by normative [expectations]''~\cite{zolyomi2023designing}. 

Tenet 3 departs from previous scholarship by emphasizing mutual benefit. Design for co-liberation asserts that neurodivergent expertise is significant not only in assistive technology for neurodivergent users but also in \textit{mainstream technologies that target both neurotypical and neurodivergent users}. Using Tenet 3, researchers might consider collaborating with neurodivergent participants to design technology that benefits both neurotypical and neurodivergent users. For example, Curtis et al. suggested ``reframing high-tech AAC as an assistive technology (AT) that \textit{all} can freely engage and leverage to interface and communicate with the evironment''~\cite{curtis2022state}. 
Likewise, Zolyomi and Snyder recently designed communication technology that ``assumes a mutual and shared responsibility for establishing, maintaining, and repairing shared frames of reference'' between neurodivergent and neurotypical users, helping to ``create needed autonomy from the pressures of broader social norms''~\cite{zolyomi2023designing}. 

Design for co-liberation also goes beyond mutual benefit to explore the ways neurodivergent expertise can transform neuronormative society. Tenet 3 differs from other strengths-based approaches to assistive technology~\cite{wobbrock2011ability, brown2023assistive, bayor2021toward, pei2019we} by shifting the target user from neurodivergent individuals to the broader society, in a similar vein as Boyd's concept of ``Celebratory Technology for Neurodiversity''~\cite{boyd2023conceptualizing}. In the same way that my partner learned from my stimming behaviors (\hyperref[stimming]{Narrative 5}), researchers might engage with Tenet 3 by studying how neurodivergent individuals' adaptive responses could benefit neurotypical individuals facing similar situations. needs''~\cite{rauchberg2022imagining}. In fact, Shew argues that disabled people are uniquely positioned to handle the uncertainty that humanity will face, as ``disabled people have expertise in navigating worlds not built for us---worlds that are often actively hostile to us,''~\cite{shew2023against}. 

\section{Author Reflections and Limitations} \label{sec:limitations}

\begin{quote}
    \textit{``I do not beg pity for my child, but I beg society to give my child the sympathy, patience, love and
opportunities to come into society and with people.''} - Anonymous parent of an autistic child from Hanoi, Vietnam~\cite{vu2014understanding}
\end{quote}
In a world where neurodivergent and disabled people are seen as broken and inferior, writing this paper has been a radical act of self-love. Back in mid-2023 when I first discovered HCI neurodivergence research, I struggled to relate to Critical Disability scholarship~\cite{boyd2024neuro}, taking weeks to parse a single paper. I was frustrated by how difficult it was for me to understand the theories and derive practical applications for my life---and I had a background in computer science, accessibility, and sociology. I could understand why there were tensions between accessibility theory and practice~\cite{mcdonnell2023tackling, spiel2020nothing, kirkham2021disability}. At the same time, I, too, felt incredibly hurt and misunderstood by the assumptions in existing technologies for neurodivergent users. I could understand why neurodivergent and disabled scholars did not want to adjust their communications to accommodate non-disabled scholars, at a further cost to themselves~\cite{williams2019prefigurative, ymous2020just}. But I firmly believe that compassion---``fueled by understanding and accepting that we're all made of strength and struggle, [that] no one is immune to pain or suffering''~\cite{brown2021atlas}---can heal divided societies~\cite{gopin2022compassionate}. So I wrote this paper---trudging through layers of self-doubt and internalized shame---as an act of love for humanity, to invite disabled and non-disabled people to come together and see that ``we are more alike, my friends, than we are unalike''~\cite{Angelou}.

Despite the focus on intersectionality, this paper, and the work described within, has been completed by a single person and is therefore not representative of the neurodivergent or disability communities. To be able to obtain a formal diagnosis, regardless of the challenges I faced, is an immense privilege in and of itself. The experiences included in this paper are shaped by my current position as a cisgender, Southeast Asian American, educated, upper class, neuroqueer\footnote{Termed by neurodivergent scholars M. Remi Yergeau, Athena Lynn Michaels-Dillon, and Nick Walker, neuroqueer is an identity that describes an individual who engages in ``the practice of queering (subverting, defying, disrupting, liberating oneself from) neuronormativity and heteronormativity simultaneously''~\cite{Walker_2021}.} woman living in the United States. I have experience building assistive technology for vision, motor, cognitive, and mental health disabilities. Additionally, my understanding of my own neurodivergence is largely dominated by autism and ADHD (with some acknowledgment of C-PTSD and mental health conditions). I encourage readers to surface and engage deeper with other intersectional voices missing from academia, including the ones mentioned in this paper (Section \ref{underheard-voices}).

\section{Conclusion}
In this paper, I provide a critical autoethnographic analysis of the ways my intersectional identities influenced my neurodivergent lived experiences, producing three tenets for reconceptualizing neurodivergence to be more inclusive of intersectionality. First, neurodivergence is a functional difference that can be adaptive, depending on context. Second, neurodivergent disability is a moment of friction enacted by an individual's abilities and their environment. Lastly, neurodivergence accessibility learns from neurodivergent traits to mutually benefit both neurotypical and neurodivergent people. In my discussion of the tenets, I demonstrate how the Data Feminism principles~\cite{d2023data} can be used as accessible and practical guidelines for engaging with intersectionality in HCI. I conclude with three recommendations for how accessibility researchers can operationalize these tenets in future work on neurodivergence: (1) investigate the diverse contexts in which neurodivergent traits are adaptive, (2) actively challenge behavioral norms when building technology, and (3) design for co-liberation. In doing so, I hope to progress the conversation between accessibility theorists and practitioners from ``Where are we now?'' and ``How did we get here?'' to \textit{``What do we want our relationship to be going forward, and what do we need to do, even if we still disagree, to create that future?''}~\cite{brown2017braving}.

\begin{acks}
I would like to express my deepest gratitude to Jina Suh for nurturing my potential as a researcher, for encouraging me to write this paper, and for believing that my ideas were worthy of being shared, even though I was a non-academic. I also thank John Porter, Avery Mack, Stacy Hsueh, Elaine Short, Mary Czerwinski, Pranav Khadpe, Fernanda De La Torre Romo, Mara Kirdani-Ryan, Gonzalo Ramos, Jason Grieves, and Tiffany Do for their guidance and support along my research journey. Thanks to Katta Spiel, my ASSETS  mentor, as well as the anonymous reviewers whose valuable insights and feedback helped shape this paper. Additionally, I am grateful to the members of the academic community who made time to engage with my initial ideas: Andrew Begel, Rua Williams, Barbara Tannenbaum, LouAnne Boyd, Gabriela Marcu, Meryl Alper, Michael Ann DeVito, Kristy Johnson, Leanne Chukowskie, Jill Lehman, Ed Cutrell, Varun Mishra, and Kentaro Toyama. I would also like to thank Orion Kelly for creating content that helped me understand and educate others about my neurodivergence. Thanks to Lucy Adams for coaching me on how to undertake such a big project by myself. Lastly, I would like to offer my appreciation to Trey Aguirre, Elizabeth Zhao, and Jeannie Le for their enduring patience, understanding, and compassion as I pursue my research aspirations. 
\end{acks}

\balance
\bibliographystyle{ACM-Reference-Format}
\bibliography{bibliography}

\appendix
\section{Data Feminism Principles} \label{data-feminism-principles}
Below is a full-text version of the Data Feminism principles, copied verbatim.

\begin{quote}
\textbf{Principle 1: Examine power.} Data feminism begins by analyzing how power operates in the world.

\textbf{Principle 2: Challenge power.} Data feminism commits to challenging unequal power structures and working toward justice.

\textbf{Principle 3: Elevate emotion and embodiment.} Data feminism teaches us to value multiple forms of knowledge, including the knowledge that comes from people as living, feeling bodies in the world.

\textbf{Principle 4: Rethink binaries and hierarchies.} Data feminism requires us to challenge the gender binary, along with other systems of counting and classification that perpetuate oppression.

\textbf{Principle 5: Embrace pluralism.} Data feminism insists that the most complete knowledge comes from synthesizing multiple perspectives, with priority given to local, Indigenous, and experiential ways of knowing.

\textbf{Principle 6: Consider context.} Data feminism asserts that data are not neutral or objective. They are the products of unequal social relations, and this context is essential for conducting accurate, ethical analysis.

\textbf{Principle 7: Make labor visible.} The work of data science, like all work in the world, is the work of many hands. Data Feminism makes this labor visible so that it can be recognized.~\cite{d2023data}
\end{quote}

\end{document}